\documentclass[12pt]{article}
\usepackage{epsfig}
\usepackage{makeidx}
\usepackage{latexsym}
\usepackage{amsfonts}
\usepackage{amssymb}
\usepackage{euscript}
\usepackage{eufrak}
\usepackage{color}
\usepackage{graphics}
\usepackage{graphicx}

\begin{document}

\title{\bf Four-wave mixing instabilities in tapered and photonic crystal fibers}

\author{F. Biancalana, D.V. Skryabin, and A. Ortigosa-Blanch \\ {\em Department of Physics, University of
Bath}
\\ {\em Bath BA2 7AY, United Kingdom}}

\maketitle

\begin{abstract}
We present an analytical study of four-wave mixing instabilities
in tapered fibers and photonic crystal fibers. Our approach avoids
the use of Taylor expansion for the linear susceptibility and the
slowly-varying envelope approximation. This allows us to describe
the generation of sidebands strongly detuned from the pump wave
with simultaneous account for the entire dispersion characteristic
of a fiber, which is found to be important for describing properly
the key role of the parametric instabilities in the supercontinuum
generation in these kind of fibers.
\end{abstract}

\noindent                 % \section*{\small{} NOT ALLOWED IN OPTICS LETTERS

\section*{\small{INTRODUCTION}}
Four-wave mixing (FWM) is one of the most fundamental processes in
optics, consisting in the generation of a pair of Stokes and
anti-Stokes photons out of two strong pump photons
\cite{stolenbjorkholm,aso}. FWM can be observed in a wide range of
materials with Kerr nonlinearity including optical fibers
\cite{agrawal1}. The efficiency of FWM (which is a coherent
process) strongly depends on the so-called phase matching
conditions imposed on the wave-vectors of the pump and generated
waves \cite{agrawal1}. In a conventional single-cladding optical
waveguide, the phase matching conditions are satisfied only in the
vicinity of the zero-dispersion point, as it has been demonstrated
for dispersion-shifted fibers and for standard telecommunication
fibers \cite{washio,serkland}, even if there are several
techniques (for example using modal birefringence) which can
extend the phase-matched frequency-range \cite{stolenboschlin}.

Several experimental results on observation of four-wave mixing in
photonic crystal fibers (PCFs) \cite{knightbirksrussell} and
tapered fibers (TFs) \cite{tapered} have been published over the
last several years \cite{coen,dudley,kumar}. These relatively new
kind of fibers represent an ideal system for investigating the
optical nonlinearities in fused-silica, because of their unique
dispersive and nonlinear properties (see for example
\cite{zheltikov,broeng}). In particular, the enhanced nonlinearity
due to the smallness of the effective core area can increase
dramatically every nonlinear effect.

The phase matching conditions for these fibers are found to be
quantitatively different with respect to ordinary fibers: phase
matching in PCFs and TFs can be achieved for a long range of pump
wavelengths, because the strong waveguide contribution to the
overall dispersion permits a compensation of the material
dispersion for a broad window of frequencies. On the other hand,
the improved nonlinearity can generate a nonlinear coefficient
which can further improve the compensation in the phase.

Moreover, the dispersion characteristics of TFs \cite{tapered} can
be very similar to the ones found in PCFs. This is because in a
typical index-guiding PCF most of the light is guided in the tiny
silica core surrounded by a periodic structure of large air filled
holes separated by thin membranes, which makes it very similar to
TFs with guiding due to total internal reflection at the interface
between silica and air \cite{tapered}, and in fact the eigenvalue
equation used to find the group velocity dispersion (GVD) in TFs
can describe with a good approximation the propagation of light in
PCFs as well \cite{snyder}, provided that we substitute the
complex cladding structure of PCF with an effective refractive
index \cite{knightbirksrusselldesandro}. Thus, because of the fact
that the GVD profile is qualitatively similar in these fibers,
they can have analogous properties in the dynamical behavior of
light propagation, as we shall see in the following.

Fig. 1 shows numerically calculated examples of the frequency
dependence of the dispersion parameter
$D=-(2\pi\omega^{2}/c)\partial^{2}_{\omega}\beta$, where $\beta$
is the propagation constant of the fundamental mode
\cite{agrawal1}, for a PCF with an elliptical core surrounded by
$4$ periods of air holes with an average diameter of $0.84\ \mu m$
and a pitch of $\Lambda=1.46\ \mu m$, and a TF with $1\ \mu m$ of
core diameter, used in recent experiments on solitonic fission and
supercontinuum generation (SCG) carried out by our group
\cite{tapered,arturo,ultimatapered}. Both dispersion
characteristics show two zero dispersion (ZD) points limiting a
central region of anomalous group velocity dispersion (GVD), which
is a typical situation for these fibers.

In several important studies FWM in PCFs has been identified as
one of the primary mechanisms contributing to generation of
optical supercontinuum by ultrashort pulse propagation, at least
in the first stage of the spatial propagation
\cite{coen,dudley,herrmann3}.

However, the analytical calculations complementing these results
were either based on the standard analysis of some improved
versions of the nonlinear Schroedinger equations (NLSE)
\cite{millotetal} or on the direct analysis of the phase matching
conditions \cite{coen,dudley,kumar,song}.

In this paper we investigate theoretically the role of parametric
instabilities in PCFs and TFs, and we establish a deep link
between these instabilities and the broad spectrum obtained in
SCG, stimulated by recent experimental works (\cite{coen,dudley})
in which the primary mechanism of spectral broadening is
identified as the combined action of stimulated Raman scattering
and parametric FWM. We shall find out that the unusual GVD profile
of these kind of fibers will play a key role in determining the
instability regions, and the analysis of these instabilities
cannot be rigorously treated with a Taylor expansion around the
carrier frequency of the pulse because of its flatness and
curvature. Moreover, we propose a mechanism for exciting a broad
window of frequencies, based on the merging between different
instability regions.

\section*{\small{MODEL EQUATION AND STABILITY ANALYSIS}}
The formal mathematical approach to the FWM problem is to study
the stability of the strong pump wave in the presence of  weak
perturbations with different frequencies. The typical instability
of the pump wave occurring in optical fibers is the so-called
modulational instability (MI) \cite{agrawal1}, which is a
particular case of the more general phenomenon of parametric FWM.
More specifically, MI can be interpreted as a four-photon process
driven by the self-phase modulation (SPM), and it is a general
feature of wave propagation in dispersive nonlinear media. In
optical fibers MI is due to the conservative interaction between
nonlinearity and linear dispersion, which leads to a modulation of
amplitude and phase of a continuous wave (CW) in presence of
quantum noise or any other weak perturbation. Traditionally the
theoretical approach to describe MI in fibers is performed using
the nonlinear Schroedinger equation, that is the simplest equation
which takes into account the interplay between dispersion and
nonlinearity. NLSE in the context of fiber optics is derived from
Maxwell's equations under the assumption of the weakly-guiding
weakly-nonlinear approximations and using a reductive perturbation
procedure consisting  of the spatio-temporal slowly-varying
envelope approximation (SVEA) and Taylor expansion of the linear
susceptibility: $\chi^{(1)}$ \cite{agrawal1}. The expansion of the
latter is often done up to second order, which corresponds to the
quadratic frequency dependence of the mode propagation constant
$\beta$: $\beta\sim\omega^2$. If GVD at the carrier frequency
$\omega$ is relatively small, then higher order terms in the
expansion of $\chi^{(1)}$ become relevant to describe propagation
of waves in fibers in general and MI in particular, see e.g.
\cite{potasek}.

However, for some fibers with complex dispersion profile and/or in
cases, when FWM results in generation of frequencies far detuned
from the pump, the NLSE-based approach and in particular the
Taylor expansion for $\chi^{(1)}$ become inapplicable and new
approaches have to be developed. One of them is to introduce a
system of coupled NLSEs for pump, Stokes and anti-Stokes waves
\cite{agrawal1}.

Another more sophisticated method is to avoid the Taylor expansion
of $\chi^{(1)}$ and analyze the frequency mixing using a single
equation, see \cite{blowwood} and \cite{agrawal2}. In
\cite{blowwood} this was mainly done in order to analyze the
influence of the Raman effect on MI, and in \cite{agrawal2} it was
applied to describe FWM in dispersion-flattened fibers. Our goal
here is to further develop the theories reported in
\cite{blowwood,agrawal2} and consider PCFs and TFs as examples,
thereby providing a theoretical underpinning for recent
experimental results \cite{coen,dudley,kumar,millotetal}.

We begin our analysis from the nonlinear wave equation derived
directly from Maxwell's equations:
\begin{equation}
\nabla^2\mathbf E-\mathbf\nabla(\mathbf\nabla\cdot\mathbf E)
-\frac{1}{c^2}\partial_t^2(\mathbf E-\mathbf P_L-\mathbf
P_{NL})=0, \label{eq1}\end{equation} where the linear polarization
is defined as
\begin{equation}
\nonumber  \mathbf{P}_{L}=\int\chi^{(1)}(t-t',x,y)\mathbf
E(t')dt'.\end{equation} $\mathbf{E}$ is the electric field, and
$\chi^{(1)}$ is the linear susceptibility of the dielectric medium
(in our case silica glass), which  depends on time and the
transverse coordinates $x,y$. $\mathbf P_{NL}$ is assumed to have
a simple Kerr form: $\mathbf{P}_{NL}=\chi_{3}|\mathbf
E|^{2}\mathbf E$. Even though the Raman effect (i.e., a delayed
response of the nonlinear susceptibility) can be incorporated into
the theory developed below, analysis of details of its influence
on FWM is well known and goes beyond the objectives of this paper
\cite{blowwood,golovchenko}.

Our next step is to reduce Eq. (\ref{eq1}) to an equation in $z$
and $t$ only. In order to achieve this we first  transform Eq.
(\ref{eq1}) from time to frequency domain, using Fourier transform
$\cal F$, and then separate transverse and longitudinal degrees of
freedom through the approximate factorization
\begin{equation}
\label{eq:fact} {\cal F}\mathbf E(x,y,z,t)=\hat{\mathbf
E}(x,y,z,\omega) \simeq \mathbf F(x,y,\omega)\tilde E(z,\omega).
\end{equation}
$\mathbf F$ is  the fundamental eigenmode of the linear waveguide
having propagation constant $\beta(\omega)$. $\beta(\omega)$
incorporates both material and waveguide contributions into the
overall fiber dispersion. The factorization becomes possible due
to the weakness of the energy transfer from the fundamental mode
to the higher order modes \cite{herrmann1,herrmann2}.

It can be shown that after using Eq. (\ref{eq:fact}) the dynamics
of the inverse Fourier transform ${\cal F}^{-1}\tilde E=\bar
E(z,t)$ of the amplitude $\tilde E$ is governed by the equation
\begin{equation}
c^2\partial_z^2\bar{E}-
\partial_t^2\bar{E}=\partial_t^2\left[
\int_{-\infty}^{t}\chi_{eff}^{(1)}(t-t')\bar{E}(t')dt'+\bar{\chi}_3|\bar{E}|^2\bar{E}\right],
\label{eq2}
\end{equation}
where $\chi^{(1)}_{eff}$ is an effective linear susceptibility of
the fiber, with its Fourier transform given by
\begin{equation}
\hat\chi^{(1)}_{eff}(\omega)={\cal
F}\chi^{(1)}_{eff}=\beta^2c^2/\omega^2-1
\end{equation} and
\begin{equation}
\bar{\chi}_3=\frac{\int{|\mathbf{F}|^{4}dS}}{\int{|\mathbf{F}|^{2}}dS}\chi_{3},
\end{equation}
where $S$ is the fiber area \cite{snyder}. $\bar{\chi}_{3}$, with
good accuracy, can be considered as a frequency independent
coefficient, which follows from the weak frequency dependence of
${\mathbf F}$ and $\chi_{3}$ \cite{agrawal1}.

Thus we have derived a version of the wave-equation with delayed
linear and instantaneous nonlinear responses. Note that Eq.
(\ref{eq2}) is an equation for the field, and not for the
envelope. Expanding $\hat{\chi}^{(1)}_{eff}$ into Taylor series
for small frequency detunings and making standard SVEA (in space
and in time), one can easily reduce (\ref{eq2}) to the
conventional NLSE.
%Note that neither \cite{blowwood} nor \cite{agrawal1} studied
%Eq. (\ref{eq2}) in its unreduced form.

We assume now that our fiber is pumped by a monochromatic
continuous wave (CW)
\begin{equation}
\bar{E}=E_{0}e^{i(kz-\omega t)}+c.c.,\label{eq3}
\end{equation}
where $E_{0}$ is a constant amplitude,  $k$ and $\omega$ are the
wave-number and the frequency of the pump wave. It is easy to see
that Eq. (\ref{eq3}) solves Eq. (\ref{eq2}) and represents a
steady-state solution, provided that the following dispersion
relation is satisfied:
\begin{equation}
k^2(\omega,|E_0|^2)=\frac{\omega^2}{c^2}\left[
1+\hat{\chi}_{eff}^{(1)}(\omega)+\bar{\chi}_3|E_{0}|^{2}\right].
\label{eq4}\end{equation} To find the frequencies generated within
the fiber as a result of the destabilization of the pump wave, we
perturb solution (\ref{eq3}) with small complex perturbations
$\epsilon(z,t)$:
\begin{equation}E=
\left[E_{0}+\epsilon(z,t) \right]e^{i(kz-\omega t)}+c.c.
\label{eq5}\end{equation} After substitution of (\ref{eq5}) into
the governing equation (\ref{eq2}) we disregard all the terms
nonlinear in $\epsilon$. Then we assume that the complex
perturbation has the following simple form, which excites both the
Stokes and the anti-Stokes bands:
\begin{equation} \epsilon(t)= A\exp(i\kappa z-i\delta
t)+B^{*}\exp(-i\kappa^*z+i\delta t),
\end{equation} and derive a linear eigenvalue problem for the vector
$(A,B)^T$ with an eigenvalue $\kappa$. It is possible to prove
that the solvability condition for $A$ and $B$ requires that the
following fourth-order polynomial equation in $\kappa$ is
satisfied:
\begin{eqnarray}
\nonumber
&&\kappa^{4}-(k^{2}_++k^{2}_-+2k^2)\kappa^{2}+2k(k^{2}_+-k^{2}_-)\kappa\\
&&+(k^2_+-k^2)(k^2_--k^2)-{\bar\chi_3^2\over
c^4}E_0^4(\omega+\delta)^2(\omega-\delta)^2=0,
\label{eq6}\end{eqnarray} where $k_{\pm}$ are the effective wave
numbers of the Stokes and anti-Stokes waves, which include
nonlinear corrections: $k_{\pm}\equiv
k(\omega\pm\delta,2|E_0|^2)$.

Solution (\ref{eq3}) becomes unstable provided that Eq.
(\ref{eq6}) has at least one root such that $Im(\kappa)<0$. As
usual, the instability gain is given by
\begin{equation}
\label{gain} g(\delta)\equiv 2|Im(\kappa)|.
\end{equation}
In the next section we explore the numerical solutions of Eq.
(\ref{eq6}).

\section*{\small{FOUR-WAVE MIXING INSTABILITIES}}
We have scanned numerically the dependence of the imaginary parts
of all four roots of (\ref{eq6}) from $\delta$ for different
values of pump frequency $\omega$ and pump power choosing the
dispersion profiles shown in Fig. 1. Typical instability-gain
profiles that we have obtained are shown in Fig. 2. We have found
that there are two - symmetric with respect to the pump frequency
- {\em pairs} of instability peaks in the spectrum of the
perturbation. All these instability peaks are associated with the
same root $\kappa=\kappa_{cr}$ of Eq. (\ref{eq6}). The regions
nearest to the pump frequency can be traced back to the
conventional MI, known from the NLSE, and the second peaks are due
to the FWM process with far detuned frequencies, as we shall see
below.

Condition for both of these instabilities to exist is that the
pump frequency should lay between the two ZD points or slightly
outside this interval, see Figs. 2 and 3.

In Fig. 2(a,c) the instability gain for our PCF and TF is shown,
pumping in anomalous dispersion, with a frequency in between the
two ZD points, $\omega_{ZD1}$ and $\omega_{ZD2}$. The GVD profiles
are also depicted [curve (1)], in order to show that the magnitude
of the instability detunings is of the order of the anomalous
regime width, $|\omega_{ZD2}-\omega_{ZD1}|$. Moving slightly
towards normal dispersion [Fig. 2(b,d)], a gap between the central
frequency and MI appears. This has been observed in several
numerical simulations, for example in Coen {\em et al.}
\cite{coen} and in Dudley {\em et al.} \cite{dudley}. However,
increasing the pump power one can observe a merging phenomenon
between the two kinds of instabilities [curve (3) in Fig. 2],
which is even more evident in normal dispersion, due to the fact
that in this regime a slight change in the pump frequency towards
the deep normal dispersion region corresponds to a large change in
the position of MI, while the FWM peaks are almost fixed.

There is a simple theoretical explanation for having instabilities
even when $\beta''>0$, using the conventional phase matching
condition
\begin{equation}
k(\omega+\delta,2|E_0|^2)+k(\omega-\delta,2|E_0|^2)-2k(\omega,|E_0|^2)=0.
\label{eq7}
\end{equation}
Expanding $k$ in Taylor series, knowing that
$\beta(\omega)\equiv(\omega/c)\left[
1+\hat{\chi}_{eff}^{(1)}(\omega)\right]^{1/2}$, and considering
the detuning $\delta$ as a small parameter, one obtains that there
is a cancellation between the odd order derivatives in the
expansion \cite{potasek}, and it can be easily proved that Eq.
(\ref{eq7}) can be written as
\begin{equation}
\label{eq:generdispersion}
\frac{1}{2}\beta''\delta^{2}+\frac{1}{24}\beta''''\delta^{4}+...\equiv\frac{1}{2}\delta^{2}
\int_{0}^{1}\left[
\beta''(\omega+\xi\delta)+\beta''(\omega-\xi\delta)
\right](1-\xi)d\xi=-\bar{\chi}_{3}|E_{0}|^{2}
\end{equation}
Even if the pump is injected in a condition of small normal
dispersion ($\beta''>0$), there can be a contribution due to the
even order terms present in Eq. (\ref{eq:generdispersion}). So in
principle the integral in Eq. (\ref{eq:generdispersion}) can be
negative, making possible the generation of instability in the
normal dispersion regime. This simple analysis of course is valid
for small frequency detunings $\delta$, but our equation
(\ref{eq6}) does not have this limitation.

One point we want to stress is that the dispersion inserted in Eq.
(\ref{eq6}) has to be considered globally, and not locally, in the
sense that for a given pump frequency the entire shape of the GVD
curve between and outside the two ZD points has significant
influence on the FWM instabilities. Thus, Taylor expansion around
a single ZD point is not adequate in this case if one wants to
capture all the features of the process.

Finding zeros of the function $\partial_{\delta}Im(\kappa_{cr})$,
which correspond to the positions of the maximal gains occurring
for $\delta=\delta_{cr}$, we calculate a diagram showing
frequencies $\omega_{AS,S}=\omega\pm\delta_{cr}$ of the generated
waves as functions of $\omega$, see Fig. 3. Fig. 3(a) shows the
position of Stokes and anti-Stokes waves for every pump frequency
and for three different powers in the case of PCF, while Fig. 3(b)
is for our TF. Dashed curves correspond to the two ZD points. It
is possible to see the merging process when increasing the pump
power (respect to a fixed pump frequency) or moving the pump
frequency (respect to fixed pump power).

Note that increasing the power the distance between the MI peaks
increases considerably, while the position of FWM peaks remains
approximately the same. Moreover, one of the two ZD points (in our
case the blue-shifted one, but of course it depends on the
particular GVD that we choose) is more "efficient", in the sense
that the frequency range excited by the instabilities is larger
respect to the other ZD point. An important difference (due to the
absence of a Taylor expansion in our equations) between Fig. 3 and
other figures present in the literature (for example Ref.
\cite{coen,dudley}) is that our curve is closed, and then the
unstable region in the normal dispersion is finite. A simple
prediction which follows from this consideration is that it should
exist a pump frequency in the normal regime for which all these
instabilities are strongly suppressed and completely disappear.

To demonstrate that instabilities found here are essentially FWM
instabilities we also show the frequencies of the generated waves
calculated using the standard phase matching condition
(\ref{eq7}). One can see remarkable agreement between results
obtained using the exact solution of Eq. (\ref{eq6}) and condition
(\ref{eq7}), confirming the fact that starting with a forward pump
wave, we can neglect the influence of the back-propagating wave in
the perturbation analysis. Note that the phase matching diagrams
presented in \cite{coen,dudley} in the vicinity of the ZD point
with largest frequency capture only a pair of instability peaks
associated with conventional MI.

Existence of the far detuned FWM branches shows that contribution
of the FWM processes to the generation of optical supercontinuum
can be in some cases even stronger than it has been so far
anticipated. Also the initial stage of supercontinuum generation
can be attributed to the threshold mechanism of merging between
the FWM peaks and the MI peaks \cite{herrmann3}, clearly visible
from Fig. 2 when one increases the pump power. Through this
mechanism it is possible to excite all the frequencies between the
two FWM peaks, and it may represent an important piece towards the
understanding of SCG.

%The two typical characteristic timescales of the process can be
%deduced directly from fig 2: there is one fast timescale related
%with MI

Making qualitative comparison of our results with these
experiments is however, complicated by the silica absorption,
which is not taken into account in our theory, that is substantial
in the far infrared (over $2$ microns). Therefore one should
expect that far red detuned radiation is strongly absorbed, and
this can potentially modify instability conditions and shift the
position of the frequencies with maximal gain. Even the Raman
effect can affect the instability gain introducing an asymmetry,
i.e. a suppression of blue-shifted frequencies respect to the
red-shifted ones.

Approximating the propagation constant in Taylor series, one can
show that secondary FWM peaks appear in correspondence of the
frequency detuning
\begin{equation}
\label{eq:positionfwm}
|\delta|\simeq\left(\frac{-12\beta^{\prime\prime}(\omega)}{
\beta^{\prime\prime\prime\prime}(\omega)}\right)^{1/2}
\end{equation}
From Eq. (\ref{eq:positionfwm}) it is evident that the
second-order dispersion $\beta''$ and the fourth-order dispersion
$\beta''''$ must have different sign in order to generate the
peaks associated to FWM.

Thus pumping the fiber in the region where GVD as function of
$\omega$ is flatter, i.e.
$\beta^{\prime\prime\prime\prime}(\omega)$ is smaller, will
generate a broader spectrum. In our examples $\beta''''(\omega)$
is smaller near the blue shifted ZD point, therefore pumping in
its vicinity one can expect generation of the broadest possible
spectrum.

An important estimate, which can be found analytically, is for the
critical power required for merging of the MI and FWM peaks. It
can be shown that the power required for merging is given by
\begin{equation}
\label{eq:powmerg} E_m^2\simeq
\frac{6n_{eff}(\omega)c}{\bar{\chi}_{3}\omega}\frac{|\beta^{\prime\prime}(\omega)
|^2}{|\beta^{\prime\prime\prime\prime}(\omega)|}
\end{equation}
$E_m^2$ is less for pump frequencies close to the ZD points.
Moreover, because of the factor $1/\omega$ in this relation,
$E_m^2$ is less for the blue-shifted ZD point respect to the
red-shifted one. It follows that in order to improve the width of
the generated spectrum one needs to find a compromise between the
power required to activate the process, which depends on
$|\beta''|^{2}/|\beta''''|$, and a small value of $|\beta''''|$ at
the pump frequency, which regulates the relative distance between
the peaks involved in the mechanism. Thus flat and small GVD are
the two factors responsible for generation of broad FWM spectrum
at low powers, which agrees with experimental finding on
generation of optical supercontinuum
\cite{tapered,coen,dudley,kumar,arturo,millotetal,herrmann2}.

The plots of the FWM gain in Fig. 2 are superimposed on the plots
showing the dispersion profile of the fiber. It is clear from this
figure that dispersion varies quite dramatically through the
relevant frequency domain; moreover, the order of magnitude of the
detuning relative to the generated frequencies is comparable with
the pump frequency itself, such that $|\delta|/\omega$ can not be
considered as a small parameter. These are the two reasons which
validates the necessity of avoiding Taylor expansion for
$\chi^{(1)}$.

\section*{\small{CONCLUSIONS}}
Parametric instabilities in PCFs and TFs have been studied in the
context of a general wave equation, avoiding the usual
approximations in optics, like the Taylor expansion of the linear
susceptibility and the suppression of the contribution of backward
waves. Other regions of instability has been found, corresponding
to degenerate FWM, which can exist even in a small region of
normal dispersion. These peaks of instability are found to be
relatively close to the pump frequency in our examples of PCF and
TF, but still too far to be detected through ordinary methods of
theoretical and numerical analysis. We have evaluated analytically
the power required to merge this regions to the modulational
instability regions, and we believe that this mechanism (regulated
by the usual fourth order dispersion $\beta''''$ in the language
of the ordinary NLSE) can be of fundamental importance in the
description of supercontinuum generation in fibers which have a
GVD with two zero dispersion points.

\section*{\small{ACKNOWLEDGMENTS}} % DISATTIVARE IL NUMERO

We  acknowledge J.K. Knight, P.St.J. Russell and T.A. Birks for
several useful discussions.

%% References

%% Figures
\newpage
%%
%% Figures and tables appear after References
%% One or more pages listing all figure captions should be
%% followed by the figures, one to a page.  Multipart figures
%% (2a, 2b, 2c, ...) may be placed on the same page.
%%
%% Following the figures, all tables should be placed
%% one table per page, except for long tables, which are allowed
%% to occupy multiple pages if necessary.  Table captions
%% should appear above the table to which it refers.
%%
\section*{List of figures}

\begin{figure}[htbp]
  \begin{center}
    \leavevmode
    \epsfbox{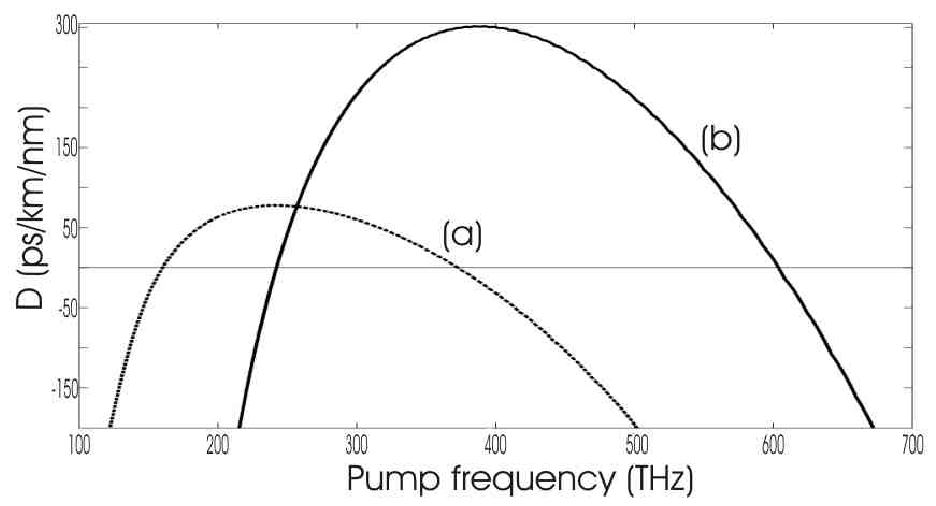}
  \end{center}
  \caption{Group-velocity dispersion (GVD) for our examples of PCF
(a) and TF (b) described in the text.}
  \label{fig:fig1}
\end{figure}

\newpage
\begin{figure}[htbp]
  \begin{center}
    \leavevmode
    \epsfbox{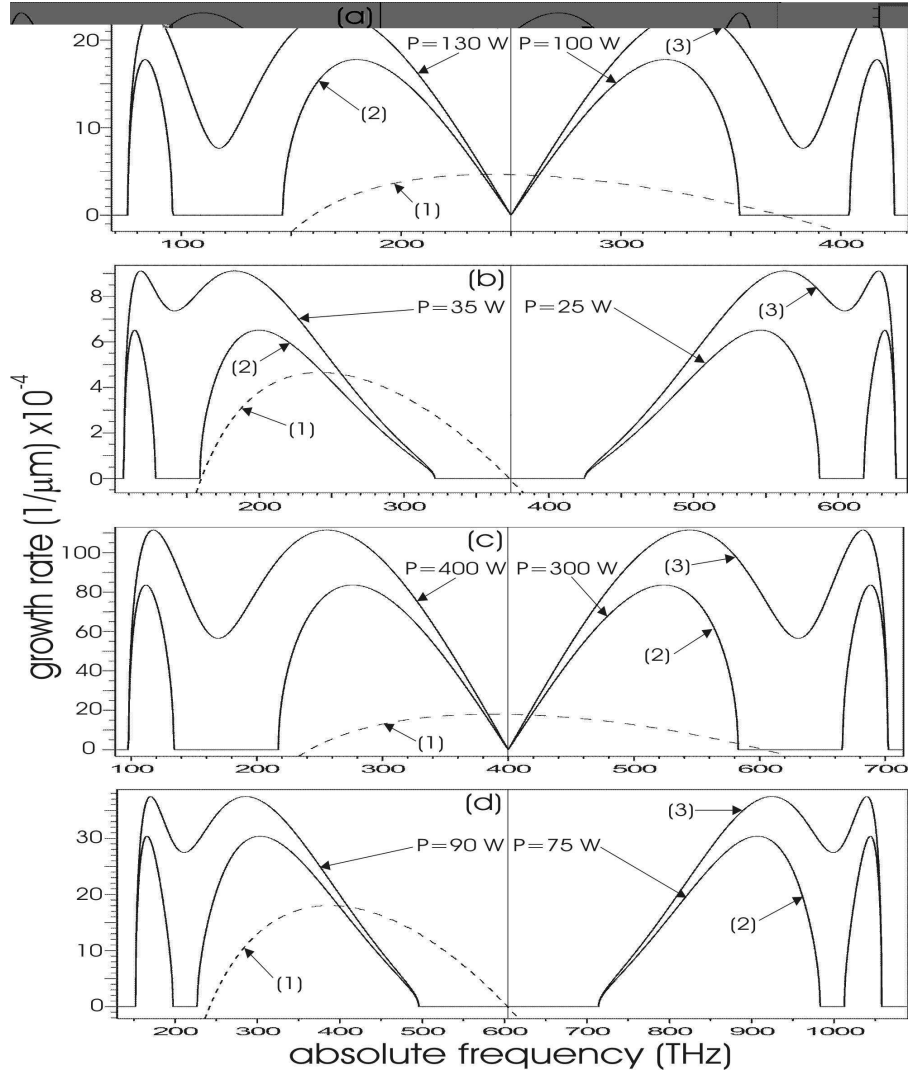}
  \end{center}
  \caption{Curves (1) show the dependence of the rescaled parameter
$D(\omega)$  on frequency for the chosen PCF and TF with $1\ \mu
m$ core diameter. To get physical values of $D$ one needs to
multiply the values for the instability gain by the factor $17$
ps/nm/km. Curves (2) and (3) show the dependence of the
instability gain $Im(\kappa_{cr})$ for  PCF (a,b) and TF  (c,d) on
frequency for different pump  powers $P=|E_0|^2$. (2) is for
$|E_0|^2<|E_m|^2$ and (3) is for $|E_0|^2>|E_m|^2$.  The central
frequency in every picture corresponds to the pump frequency.}
  \label{fig:fig2}
\end{figure}

\newpage
\begin{figure}[htbp]
  \begin{center}
    \leavevmode
    \epsfbox{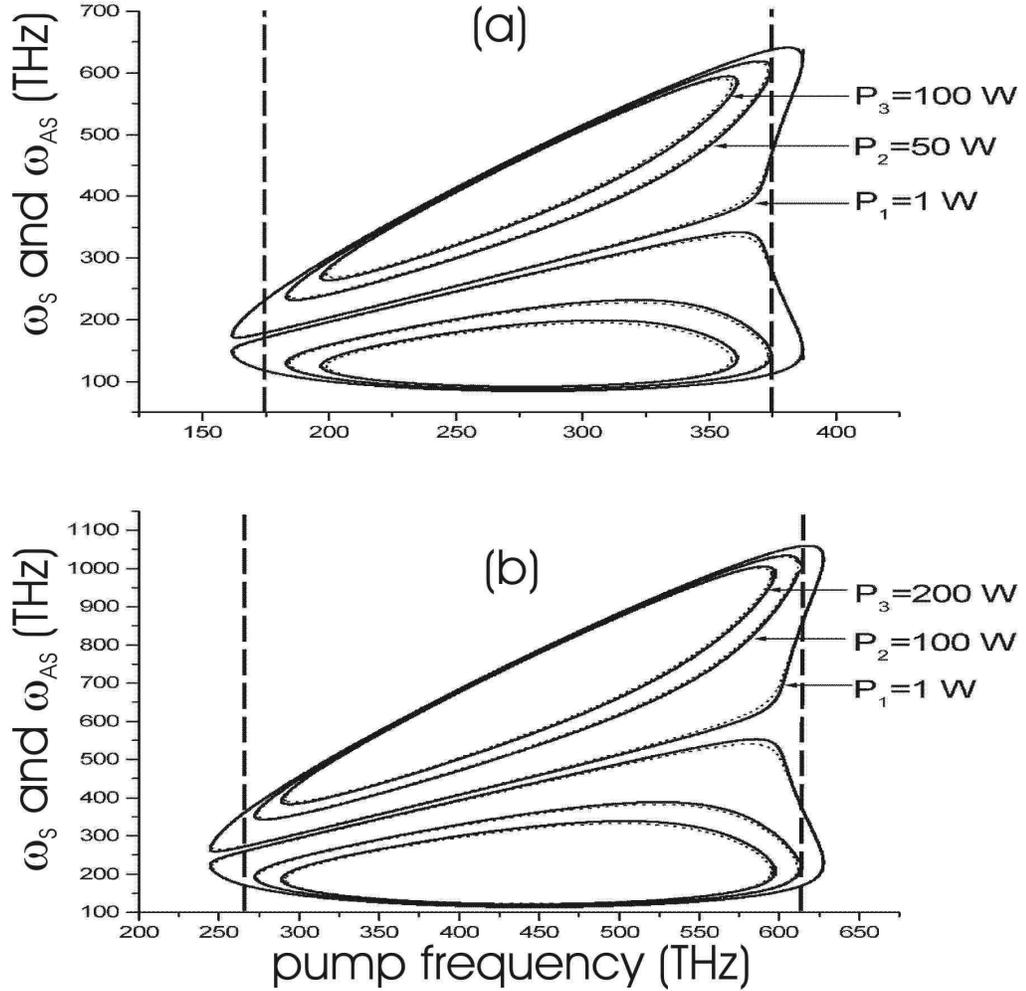}
  \end{center}
  \caption{Diagrams showing positions of the maximal instability
gain vs pump frequency calculated from Eq. (7) for different
powers $P=|E_0|^2$ in  PCF (a) and   TF (b). Dots show the same
values calculated using phase matching condition (8).  ZD points
are indicated by dashed vertical lines.}
  \label{fig:fig3}
\end{figure}

\end{document}